\documentstyle[aps,epsf]{revtex}

\begin{document}

\title{Shape coexistence in $N=14$ isotones: $^{19}$B, $^{24}$Ne and $^{28}$Si
}

\author{Y. Kanada-En'yo}

\address{Institute of Particle and Nuclear Studies,\\
High Energy Accelerator Research Organization,\\
 Tsukuba 305-0801, Japan}


\maketitle

\begin{abstract}
{Shape coexistence problems in $N=14$ isotones 
are studied with a microscopic method of antisymmetrized molecular 
dynamics. The present calculations reproduce features of deformation
in $^{28}$Si, and predict possible shape coexistence of neutron density
in neutron-rich nuclei($^{19}$B and $^{24}$Ne).
We also systematically study the structures of 
the ground and excited states, and the molecular resonances  
of $^{28}$Si. Besides the shape coexistence in the low-energy region, 
the results indicate the high-lying levels with 
$\alpha$-cluster and $^{12}$C+$^{16}$O molecular structures in $^{28}$Si,
which are consistent with the observed spin-assigned resonances.
The resonance states above the threshold energies 
are connected with the low-lying deformed states from a view point of 
molecular excitation by discussing inter-cluster wave functions.
}
\end{abstract}

\section{Introduction}

Shape coexistence phenomena have been suggested in many $sd$-shell nuclei
\cite{WOOD}.
One of the famous shape coexistence has been known
in $^{28}$Si, in which the structure of the ground band is oblate,
and an excited band 
starting from a $0^+_3$ state at 6.691 MeV is considered to be a 
prolately deformed band.
On the other hand, resonances in the $^{12}$C+$^{16}$O(C-O) 
excitation functions 
have been experimentally observed in the elastic, inelastic,
other exit channels and fusion cross sections 
\cite{JAMES,CHARLES,FROLICH,BAYE}. These molecular resonances 
lie in the excitation energy region $\sim$ 30-50 MeV in $^{28}$Si.
Another group of levels in the excitation energy region
from 18 to 30 MeV observed in $\alpha$ transfer reactions \cite{ARTEMOV}
indicates an $\alpha$-cluster structure of $^{28}$Si.

The first subject in the present paper is the shape coexistence in
$N=14$ isotones. It is a challenging problem whether or not shape 
coexistence is found in unstable $N=14$ nuclei.
The second subject is systematic study on 
structures and cluster behaviour of the
 excited states of in $^{28}$Si including the high-lying molecular resonances. 
There were many theoretical efforts for 
study on the $^{28}$Si structures.
For example, the shape coexistence in the low-energy region was studied
with mean field approaches and $7\alpha$ cluster models
\cite{BAUHOFF}, while the molecular resonances were discussed
with C-O cluster models\cite{BAYE,KATO}. However we have not comprehended 
features of those rotational bands observed
in the wide energy interval.
In the present paper, we systematically study 
the shape coexistence and the 
molecular resonant features in $^{28}$Si.
We apply a microscopic method of antisymmetrized molecular dynamics(AMD)
\cite{ENYOa,ENYObc}. 

\section{Formulation}

An AMD wave function 
is a Slater determinant of Gaussian wave packets;

\begin{eqnarray}
&\Phi_{AMD}={1 \over \sqrt{A!}}
{\cal A}\{\varphi_1,\varphi_2,\cdots,\varphi_A\},\\
&\varphi_i=\phi_{{{\bf Z}}_i}\chi_i\tau_i :\qquad
\phi_{{{\bf Z}}_i}({\bf r}_j) \propto
\exp\left 
[-\nu({\bf r}_j-{{\bf Z}_i \over \sqrt \nu})^2\right],
\label{eqn:single}
\end{eqnarray}
where the centers of Gaussians (${\bf Z}_i$) are complex variational
parameters. In the present paper, the intrinsic spin and isospin part
$\chi_i\tau_i$ is fixed to be a up/down proton or neutron.
In order to obtain an intrinsic wave function of a level,
We vary the parameters ${\bf Z}_i$ to 
minimize the energy expectation value for the
parity eigen-state projected from a Slator AMD wave function
by using a frictional cooling method.
 After the energy variation, we perform total angular momentum projection.

For the study of highly excited states, we apply the AMD method with
constraint. 
In the present calculations, we adopt the constraint
on total principal quanta ${\hat N}=n_x+n_y+n_z$ of the spherical
harmonic oscillator.
After the energy variation
(variation with the constraint ${\hat N}=W$
after parity projection) ,
we superpose the spin-parity projected 
states to diagonalize the Hamiltonian
and norm matrixes with respect to the generator coordinate $W$. 

\section{Results}
We studied the structures of $N=14$ isotones: $^{28}$Si, $^{19}$B, and 
$^{24}$Ne.
The adopted interactions in the present work are
the central force of the modified Volkov No.1\cite{TOHSAKI}  with case 3
($m=0.62$), the spin-orbit force of G3RS\cite{LS} ($u_1=-u_2=2800$ MeV)
 and the Coulomb force.
We choose an optimum width parameter
($\nu$) for the Gaussians of the single-particle wave functions
of each nucleus.

\subsection{Shape coexistence in $N=14$ isotones}
We study the normal parity states of 
$^{28}$Si, $^{19}$B and $^{24}$Ne 
with simple AMD calculations(parity projection before variation and
angular momentum projection after variation).
The ground band of $^{28}$Si is known to be oblate, 
while the excited rotational band starting from a $0^+_3$ state at
6.6914 MeV is considered to be prolate. 
Many studies with $7\alpha$-cluster models failed to describe
the large energy interval between the oblate and prolate states.
In the present AMD calculations, we obtained two local minima with 
oblate and prolate deformation.
The important point in the AMD 
results is that we can reproduce the large excitation energy of 
the prolate state because of the energy gain of the spin-orbit force.
 
\begin{figure}[ht]
\centerline{\epsfxsize=7cm\epsffile{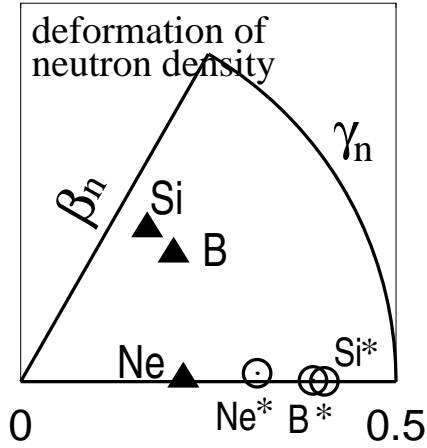}}
\caption{\label{fig:beta} Deformation parameters for neutron density.}
\end{figure}
\begin{figure}   
\centerline{\epsfxsize=12cm\epsffile{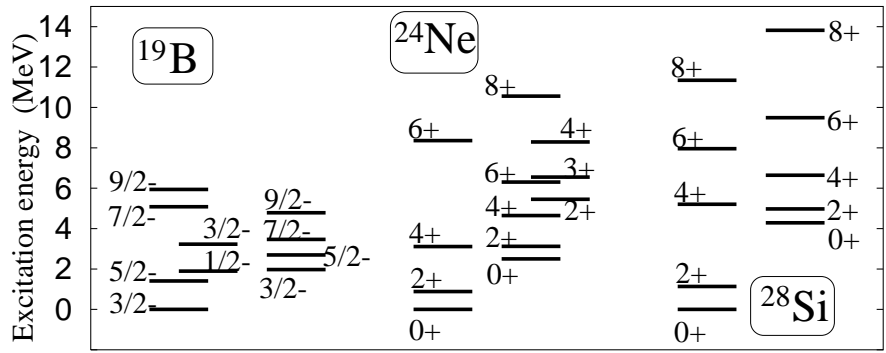}}
\caption{Excitation energies of normal parity states in 
$^{19}$B, $^{24}$Ne and $^{28}$Si calculated with simple AMD.}
\end{figure}

In the calculated results of $^{19}$B, we also found two local minima
with oblate and prolate deformation of neutron density as seen in 
Fig.\ref{fig:beta} for 
the deformation parameters.
The present calculations suggest an oblate
ground state and a prolate excited state as well as $^{28}$Si,
though excited states in $^{19}$B have not been experimentally measured yet.
In the calculated prolate states with a few MeV excitation,
it was found that a cluster structure like $^8$He+$^{11}$Li develops.
The deformation 
parameters for proton density are $\beta_p=0.3$ in the oblate state,
and $\beta_p=0.4$ in the prolate state. 
It means that the proton matter shape in $^{19}$B correlates with
the deformation of neutron density.
Therefore measurement of $Q$-moment
will be a good probe to know the deformation of $^{19}$B.
Since the calculated $Q$-moments qualitatively
depend on the adopted forces,
systematic analysis of $Q$-moments along B isotopes is rather helpful.
In the present calculations, the $Q$-moment of 
the oblate $3/2^-$ state of $^{19}$B is almost same as or even less than 
that of $^{17}$B, while that of the prolate $3/2^-$ state of $^{19}$B is 
20 \% larger than $^{17}$B. 

In $^{24}$Ne, two local minima 
were found, however,  the features of deformation are different from those in 
$^{28}$Si and $^{19}$B. That is to say, both minima have
prolate deformation of neutron density. 
The smaller prolate deformation is lowest, 
while deformation is larger in the excited band.
Since the lowest state with smaller deformation gains the spin-orbit force,  
we consider that it has an analogous neutron structure with those in the oblate
states in $^{19}$B and $^{28}$Si. The reason for 
the small prolate deformation in $^{24}$Ne is 
a trend of the prolately deformed proton structure in Ne isotopes.
Another interesting feature of shape in $^{24}$Ne is a discrepancy
between proton($\beta_p$) and neutron($\beta_n$) deformation
in the ground band: $\beta_p=0.3$ and $\beta_n=0.2$. 
The theoretical value of $B(E2;2^+_1\rightarrow 0^+_1)$
for protons(neutrons) is $B_p(E2)=37$ e$^2$fm$^4$ (
$B_n(E2)=38$ e$^2$fm$^4$).
In spite of the large number of valence neutrons,
the strength for neutrons $B_n(E2)$ is not so large, which is caused by
the small deformation of the neutron density.

\subsection{Structure of $^{28}$Si}

We studied the structure of the ground, excited and molecular resonant states
with AMD calculations with constraint on the principal quanta 
of the spherical harmonic oscillator $\hat N=W$. With a given value $W$ and each
parity, we performed energy variation 
and find some local minima, which are based on the coexisting minima
in the low-energy region.
Concerning the superposition of the obtained states,
we chose 23 AMD wave functions as base states
in the diagonalization for each parity.

\begin{figure}[ht]
\centerline{\epsfxsize=15cm\epsffile{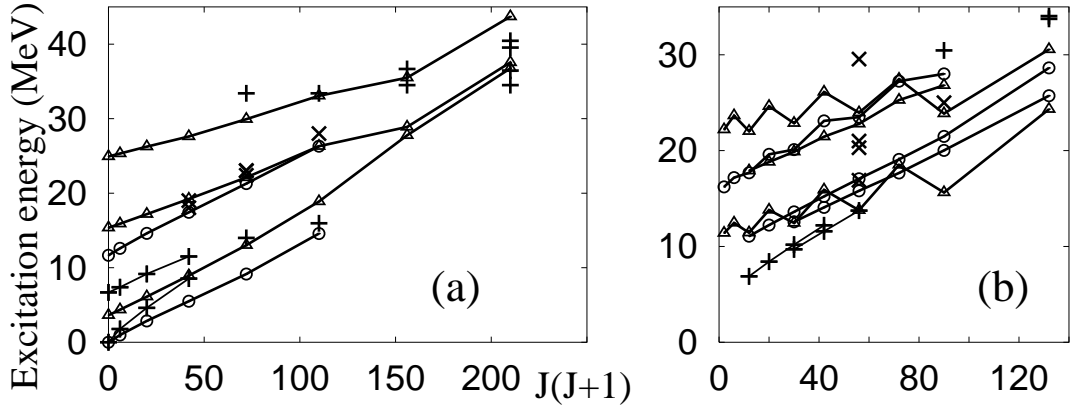}}   
\caption{\label{fig:rot} Excitation energies of $^{28}$Si.
Circles(triangles) indicate the calculated levels in the group of 
the oblate(prolate) states. Symbols ($\times$,$+$) are for the experimental
levels. 
}
\end{figure}

By analyzing the amplitudes of the base states after diagonalization, 
we can classify many of the excited states into two groups.
The first group consists of prolate states, while the second group is
a family based on oblate states. In the first group of 
prolately deformed states with positive parity, 
we found three rotational $K^\pi=0^+$ bands which are 
regarded as members of C-O molecular
bands. The calculated lowest band well corresponds to the observed 
prolate band starting at 6.69 MeV.
The second and third bands are the higher excited bands with
molecular excitation in the inter-cluster motion.
These are actually dominated by $^{12}$C and $^{16}$O
cluster states. In the extracted inter-cluster wave functions between
the $SU_3$ limit C$_{g.s}$ and O$_{g.s}$ clusters, it was found that the lowest
band has the minimum number $n_0=8$ of nodes in Pauli allowed states, while
the relative motion in the second and third bands have 9 and 10 nodes,
respectively. In the energy region around the third bands, there exist
many experimental levels with the C-O molecular resonant behaviour 
\cite{JAMES,CHARLES,FROLICH,BAYE}. 
In the negative parity states with prolate deformation, we found three-body
cluster features consisting of C-$\alpha$-C structure.
As a result, two rotational bands with $K^\pi=1^-$ consist of the axial 
asymmetric C-$\alpha$-C structure. The three-body(C-$\alpha$-C) 
cluster features have been studied 
with a model of three structure-less clusters\cite{WIEBICKE}.

In the family of oblate states, the calculated results
indicate exotic shapes in the negative parity states belonging
to low-lying 
$K^\pi=3^-$ and $5^-$ bands.
For example, a pentagon shape composes the $K^\pi=5^-$ band.
The results well correspond to the experimental energy levels and 
strength of in-band $E2$ transitions of the $K^\pi=3^-,5^-$ bands
\cite{GLATZ}.
In the positive parity states, the calculated lowest band has
oblate deformation which is consistent with the experimental feature of the 
ground band.
Above the ground band, a band with Mg-$\alpha$ cluster structure 
appears at about 10 MeV excitation energy. The moment of inertia of this band
is lower (with a rotational parameter $\hbar^2/2I\sim 130$ keV) than 
the second and third C-O molecular bands ($\hbar^2/2I\sim$ 99 and 74 keV).
This excited band is considered to be
a higher band with molecular excitation in the Mg-$\alpha$ channel 
based on the oblate ground band. 
In other words, an $\alpha$ cluster in the ground state
comes outward to make a higher nodal relative motion between the 
$\alpha$ and the Mg core. 
Around the corresponding excitation energy,
there is a group of the  experimental levels
observed in $\alpha$ transfer reactions \cite{ARTEMOV}.

\section{Summary}
We investigated the structure of $N=14$ isotones with the method of 
antisymmetrized molecular dynamics.
Shape coexistence of neutron density was theoretically suggested
in the low-energy region of $^{19}$B and $^{24}$Ne.
 
We also performed systematic study of the ground, excited states 
and molecular resonant states in $^{28}$Si.
Molecular resonant states appear in the C-O channels due to the 
excitation of the inter-cluster motion. The C-O resonances
can be connected with the lowest prolate band.
The other resonances in the Mg-$\alpha$ channel are obtained at about 10
MeV excitation. They are considered to be the
other molecular excited states based on the oblate structure in the 
ground band.
We also found the
C-$\alpha$-C behaviour in the negative parity bands.
In the high-lying states,
the molecular states with inter-cluster excitation were found in 
various kinds of channels, which are based on the
coexisting deformed states in the low energy region.
It is an interesting future problems to search for
molecular states in unstable $sd$-shell nuclei.

\section*{Acknowledgments}
The computational calculations in this work were supported by the
Supercomputer Project of High Energy Accelerator Research Organization(KEK),
and 
Institute of Physical and Chemical Research (RIKEN).

\end{document}